\documentclass[reprint,showpacs,preprintnumbers,amssymb,superscriptaddress,aps,prd,longbibliography]{revtex4-1}

\usepackage{graphicx, epsfig, amssymb} 
\usepackage{amsmath, amsfonts}
\usepackage{bm} 

\usepackage[linktocpage]{hyperref}
\usepackage[usenames]{color}
\usepackage{subfigure}
\usepackage[utf8]{inputenc}
\usepackage{tensor}
\usepackage{soul}

\def\b{\begin{equation}}
\def\e{\end{equation}}

\def\be{\begin{eqnarray}}
\def\ee{\end{eqnarray}}
\def\bf{\begin{figure}}
\def\ef{\end{figure}}

\begin{document}\title{\large Superradiance in static black hole spacetimes}

\author{Carolina L. Benone}\email{lben.carol@gmail.com}
\affiliation{Faculdade de F\'{\i}sica, Universidade Federal do Par\'a, 66075-110, Bel\'em, Par\'a, Brazil}

\author{Lu\'{\i}s C. B. Crispino}\email{crispino@ufpa.br}
\affiliation{Faculdade de F\'{\i}sica, Universidade Federal do Par\'a, 66075-110, Bel\'em, Par\'a, Brazil}

\begin{abstract}
We investigate the absorption of a massive and charged scalar field in a Reissner-Nordström background. We compare our numerical results for the absorption cross section, obtained for arbitrary frequencies, with the low- and high-frequency limits. We find, in particular, that the total absorption cross section can be negative, showing that planar scalar waves can be superradiantly amplified by black holes.

\end{abstract}

\pacs{04.70.-s, 
11.80.Et, 
04.70.Bw, 
11.80.-m, 
4.62.+v, 
4.30.Nk 
}
\date{\today}

\maketitle

\section{Introduction\label{sec:intro}}

Scattering is a matter of major interest in physics, being responsible, for instance, for the understanding of the atomic nucleus. This tool has been used to study black holes since the 60's, starting with the pioneer work by Matzner, in 1968 \cite{matzner:163}. The first investigations were performed analytically, using approximations in the limits of high and low frequencies. The development of numerical techniques performed with the aid of computers allowed the investigation of scattering and absorption problems in the full range of frequencies.

There are plenty of works on scattering by Schwarzschild black holes, including the scattering and absorption for massless \cite{matzner:163,Sanchez:1978} and massive \cite{Jung:2004yh} scalar fields, fermions \cite{Doran:2005vm,Dolan:2006vj}, electromagnetic \cite{Crispino:2009xt} and gravitational waves \cite{Vishveshwara:1970zz}. For rotating black holes there is still a lot of work to be done in this area, although some investigations have been devoted to the scattering and absorption for scalar \cite{Glampedakis:2001cx,Macedo:2013afa} and gravitational fields \cite{Dolan:2008kf} in Kerr spacetime. 

For the Reissner-Nordström spacetime there is also a considerable number of studies in the literature. Nakamura and Sato \cite{Nakamura:1976nc} studied the absorption of charged and massive scalar fields in the Reissner-Nordström spacetime, finding the reflection coefficient numerically. There are also recent works that compute the absorption and scattering cross sections of Reissner-Nordstrom black holes for massless \cite{Crispino:2009ki} and massive scalar fields \cite{Benone:2014qaa}, electromagnetic  \cite{Crispino:Oliveira:2008,Crispino:2009zza,cdho2014} and gravitational waves \cite{Oliveira:2011zz,Crispino:2015gua}. Some studies on absorption by higher-dimensional Reissner-Nordström black holes can also be found in the literature \cite{Jung:2004yn,Jung:2005mr,chm2010}.

One effect that has received a lot of attention is the superradiant scattering, which was first studied, for black holes, by Misner \cite{Misner:1972kx}. This effect is triggered when a wave, with frequency smaller than a certain critical value, is scattered by the black hole. The wave gains energy from the black hole and is scattered with more energy than it originally had. If we can trap these scattered waves in a cavity, such that they keep being scattered by the black hole, we may create a chain reaction and develop a superradiant instability, dubbed black hole bomb \cite{Press:1972zz}. Recently a comprehensive review about superradiance has been released \cite{Brito:2015oca}.

In order to have superradiance in a physical system, we must have a dissipation mechanism associated. In black hole systems, the dissipation occurs through the event horizon. We also need energy gain from the black hole for the superradiance to take place \cite{Bekenstein:1973mi}. In the Reissner-Nordström case a charged field can gain energy by stealing mass and charge from the black hole. Recently, Di Menza and Nicolas \cite{DiMenza:2014vpa} studied some aspects of superradiance in Reissner-Nordström spacetimes. Moreover, the case of superradiant instability of Reissner-Nordström black holes has been studied in Refs. \cite{Degollado:2013bha,Herdeiro:2013pia}.

Here we compute the absorption cross section for a charged and massive scalar field in a Reissner-Nordström background for the full range of frequencies, using a numerical approach. We compare our numerical results with the low- and high-frequency limits, which we obtain analytically. We also analyse the superradiant scattering implications in the partial and total absorption cross sections.

The remainder of this paper is organized as follows. In Sec. \ref{fe} we present the field equations for the massive and charged scalar field in the Reissner-Nordström spacetime. In Sec. \ref{as} we find analytical approximations for the absorption cross section in the low- and high-frequency regimes. In Sec. \ref{na} we show the results obtained numerically, comparing them with the analytical approximations obtained in Sec. \ref{as}. In Sec. \ref{fr} we present our discussions and final remarks. We assume $c=G=\hbar=1$.


%
%

\section{The field equations}
\label{fe}
The Reissner-Nordström line element can be written as
\b
ds^2 = f dt^2 - (1/f) dr^2 - r^2 (d\theta^2 + \sin^2{\theta} d \phi^2),
\e
where 
\b
f=\left(1-\frac{r_+}{r}\right)\left(1-\frac{r_-}{r}\right),
\e
with
\b
r_\pm = M \pm \sqrt{M^2-Q^2} 
\e
being the outer (event) horizon ($r_+$) and inner (Cauchy) horizon ($r_-$), where $M$ and $Q$ are the black hole mass and charge, respectively.

The Klein-Gordon equation for a charged scalar field in this geometry can be written as
\b
(\nabla_\nu - i q A_\nu)(\nabla^\nu - i q A^\nu)\Phi = \mu^2 \Phi,
\label{kge}
\e
where $\mu$ and $q$ are the mass and charge of the scalar field, respectively, and $A^\nu = (Q/r,0,0,0)$ is the vector potential of the background electromagnetic field. We can solve Eq. (\ref{kge}) by making a separation of variables, such as
\b
\Phi_{\omega l} = \frac{\psi_{\omega l}(r)}{r} P_l(\cos \theta) e^{-i\omega t},
\label{eq:separation}
\e
where $P_l(\cos \theta)$ is a Legendre polynomial, and $\psi_{\omega l}(r)$ satisfies the radial equation
\b
f\frac{d}{dr}\left(f\frac{d}{dr}\psi_{\omega l}\right) +V_\textit{eff}(r)\psi_{\omega l} = 0,
\label{fdr}
\e
with the effective potential
\b
V_\textit{eff}(r) = \left[\left(\omega-\frac{q Q}{r}\right)^2 -f\left(\mu^2+\frac{l(l+1)}{r^2}+\frac{f'}{r}\right) \right].
\e
Since we are interested in absorption and scattering properties, we assume that the scalar field is unbounded,  so that $\omega > \mu$.

We can use the tortoise coordinate $r_*$, defined by
$
dr_* / dr = f^{-1} 
\label{trc}
$, in Eq. (\ref{fdr}) and obtain a Schrödinger-like equation, namely
\b
\frac{d^2}{dr_*^2}\psi_{\omega l} +V_\textit{eff}(r) \psi_{\omega l} = 0.
\label{fdr2}
\e
Taking the asymptotic limits of Eq. (\ref{fdr2}) we find the solutions
\b
\psi_{\omega l}(r) \approx 
\left\{ 
\begin{array}{ll}
\, T_{\omega l} \, e^{-i\xi r_*}, \quad &\mbox{for $r\rightarrow r_+$},\\
 e^{-i\rho r_*} + R_{\omega l} e^{i\rho r_*}, \quad &\mbox{for $r\rightarrow \infty$},
\end{array}
\right.
\label{sol}
\e
where $\xi \equiv \omega - qQ/r_+$ and $\rho \equiv \sqrt{\omega^2-\mu^2}$. $T_{\omega l}$ and $R_{\omega l}$ are associated to the transmitted and reflected parts of the wave. By imposing the conservation of the flux, we find the relation
\b
|R_{\omega l}|^2 + \frac{\xi}{\rho}|T_{\omega l}|^2 = 1,
\label{rtf}
\e
where $|R_{\omega l}|^2$ is the reflection coefficient and $|T_{\omega l}|^2$ is the trasmission coefficient.

When $qQ<0$, $\xi>0$ and we have an incoming wave at the horizon, implying that there is no superradiance for this case. On the other hand, when $qQ>0$ and $\omega < |qQ|/r_+$ we have $\xi<0$. In this case the sign of the exponential $e^{-i\xi r_*}$ is switched and instead of an incoming wave at the horizon we have an outgoing wave. This implies in a superradiant scattering, since this outgoing wave at the horizon will reinforce the outgoing wave at infinity. 

The absorption cross section is given by 
\b
\sigma = \sum_{l=0}^\infty \sigma_l,
\e
where the partial absorption cross section $\sigma_l$ is given by
\be
\sigma_l &=&\frac{\pi}{\rho^2}(2l+1)(1-|R_{\omega l}|^2)\nonumber\\
&=& \pi\frac{\xi}{\rho^3}(2l+1)|T_{\omega l}|^2.
\ee

\section{Analytical results}
\label{as}
\subsection{Low-frequency limit}
In this subsection we compute the absorption cross section in the limit $\omega \ll 1$. In order to find an analytical solution we also assume $\mu \ll 1$ and $q Q/r_+ \ll 1$. We follow the same method used by Unruh in Ref. \cite{Unruh:1976fm}, which consists in making a separation of the parameter space in three different regions, namely:
\begin{itemize}
\item Region I, very close to the black hole event horizon, i.e., $r \rightarrow r_+$;
\item Region II, low-frequency limit, where $\omega, \mu$ and $qQ/r_+$ are much smaller than $1$;
\item Region III, far away from the black hole event horizon, i.e., $r \gg r_+$. 
\end{itemize}

We find similar equations to the ones obtained in Ref. \cite{Benone:2014qaa} for each region, which are
\b
\phi_{\omega l}(r) \approx 
\left\{ 
\begin{array}{ll}
\, A^\textit{tra} e^{-i\xi r_*}, \quad &\mbox{for Region I},\\
 \zeta \ln{\left(\frac{r-r_+}{r-r_-}\right)}+\tau, \quad &\mbox{for Region II},\\
 a\frac{F_l(\eta,\omega v r)}{r}+ b\frac{G_l(\eta,\omega v r)}{r},  \quad &\mbox{for Region III},
\end{array}
\right.
\label{slf}
\e
where $\phi_{\omega l}\equiv\psi_{\omega l}/r$; $F_l(\eta,\omega v r)$ and $G_l(\eta,\omega v r)$ are the regular and irregular spherical wave functions, respectively; $A^\textit{tra}, \zeta, \tau, a$ and $b$ are constants, and
\b
\eta = -\frac{M(\omega^2+\rho^2)}{\rho} + \frac{qQ}{2\rho}.
\e

We then make an interpolation between the regions and find the low-frequency absorption cross section, given by
\b
\sigma_\textit{lf} = \frac{\mathcal{A}}{\rho}\left(\omega - \frac{qQ}{r_+}\right),
\label{lfa}
\e
where $\mathcal{A} = 4\pi r_+^2$ is the area of the black hole. We see that for $q = 0$, Eq. (\ref{lfa}) gives us $\mathcal{A}/v$, where $v=\sqrt{1-\mu^2/\omega^2}$, which is the result obtained for the massive chargeless scalar field in a Reissner-Nordström background \cite{Benone:2014qaa}.

\subsection{High-frequency limit}
In the limit $\omega \gg \mu$ the field can be analyzed as a charged particle. Is this case the particle does not follow a geodesic, since it is subjected to the Lorentz force caused by the charged black hole. We can write the equations of motion and find the orbit equation of the particle, given by
\be
\left(\frac{d u}{d\phi}\right)^2 &=& -f(u) u^2 + (1- f(u))\frac{\mu^2}{L^2}+\frac{Q^2 u^2 q^2}{L^2}\nonumber\\
&\ &  - \frac{2 Q q E u}{L^2} + \frac{E^2-mu^2}{L^2},
\label{geo}
\ee
where $u \equiv 1/r$ and $f(u) = 1- 2M u+ Q^2 u^2$. $E$ and $L$ are the energy and the angular momentum of the particle, respectively, which, in the semiclassical limit can be associated to $E\rightarrow\omega$ and $L\rightarrow l+1/2$, respectively. 

The high-frequency absorption cross section is given by $\sigma_\textit{hf} = \pi b_c^2$, where $b_c = L_c/\sqrt{E_c^2-\mu^2}$ is the critical impact parameter and $L_c$ and $E_c$ are the critical angular momentum and critical energy, respectively. With Eq. (\ref{geo}) and its derivative we can find $b_c$ and, consequently, the high-frequency absorption cross section. We have found a lengthy closed-form solution for the critical impact parameter $b_c$, which we will not exhibit here.

\section{Numerical analysis}
\label{na}
We can solve the radial equation (\ref{fdr}) numerically, from very close to the black hole event horizon to very far away from it, using the asymptotical expressions, given by Eq. (\ref{sol}) and their derivatives. We can use the mass of the black hole as a normalization factor and fix the values of $Q, q$ and $\mu$. With the radial solution we are able to compute numerically the reflection coefficient and the absorption cross section.


Figure \ref{rcq} shows the reflection coefficient for $M\mu=0.4$, $Q/M=0.8$, $l=0$, and different choices of $Mq$. In this plot we exhibit only the part of the reflection coefficient that is greater than 1. We see that, as we increase the charge of the field, keeping its mass fixed, the maximum of the reflection coefficient grows and moves to the right, i. e., happens for bigger values of the frequency.

If we fix $M q = 1.6$ and vary $M \mu$, as in Fig. \ref{rcm}, we see that for smaller values of $M\mu$ we have greater values of the reflection coefficient. We also notice a modification in the shape of the reflection coefficient as we vary the parameter $M\mu$.

In Fig. \ref{tac} we compare the high-frequency limit of the absorption cross section with the numerical results for a fixed black hole charge and different choices of the scalar field charge. We observe that (i) the numerical results oscillate around the analytical approximation, even for intermediate values of the frequency, and (ii) as we increase the charge of the field the absorption cross section decreases. We can also observe from Fig. \ref{tac} that, depending on the charge of the field, we can have a finite value for the absorption cross section in the limit $\omega/\mu \rightarrow 1$, contrasting with the chargeless massive field case, in which we always have a divergent absorption cross section in this limit. 

Figure \ref{tacm} presents the total absorption cross section for different choices of the mass coupling $M\mu$.  We see that, for fixed values of $Q/M$ and $Mq$, as we increase the mass coupling the total absorption cross section also increases.

In Fig. \ref{pacl} we compare our numerical results for the $l=0$ case with the low-frequency absorption cross section $\sigma_\textit{lf}$, given by Eq. (\ref{lfa}). We see that for very small mass couplings $M\mu$, $\sigma_\textit{lf}$ gives a good approximation for the absorption cross section, but as we increase the mass coupling, the approximation gets worse, once that the numerical results go to infinity faster than the analytical ones.

In Fig. \ref{tpa2} we show the partial and total absorption cross sections as function of the frequency for $M\mu=0.4, Q/M=0.4$, and $M q=-0.4$. In this case $Qq<0$, therefore we do not have superradiance. Moreover, in the low frequency limit the total absorption cross section goes to infinity.

In Fig. \ref{tpa} we present the partial and total absorption cross sections for $M\mu=0.4, Q/M=0.8$ and $M q=1.6$. In the inset we can observe that for $1<\omega/\mu \lesssim 2$ the total and partial absorption cross sections for $l=0$ and $l=1$ have negative values, implying that, in this case, we have superradiant scattering.

\bf
\includegraphics[width=\columnwidth]{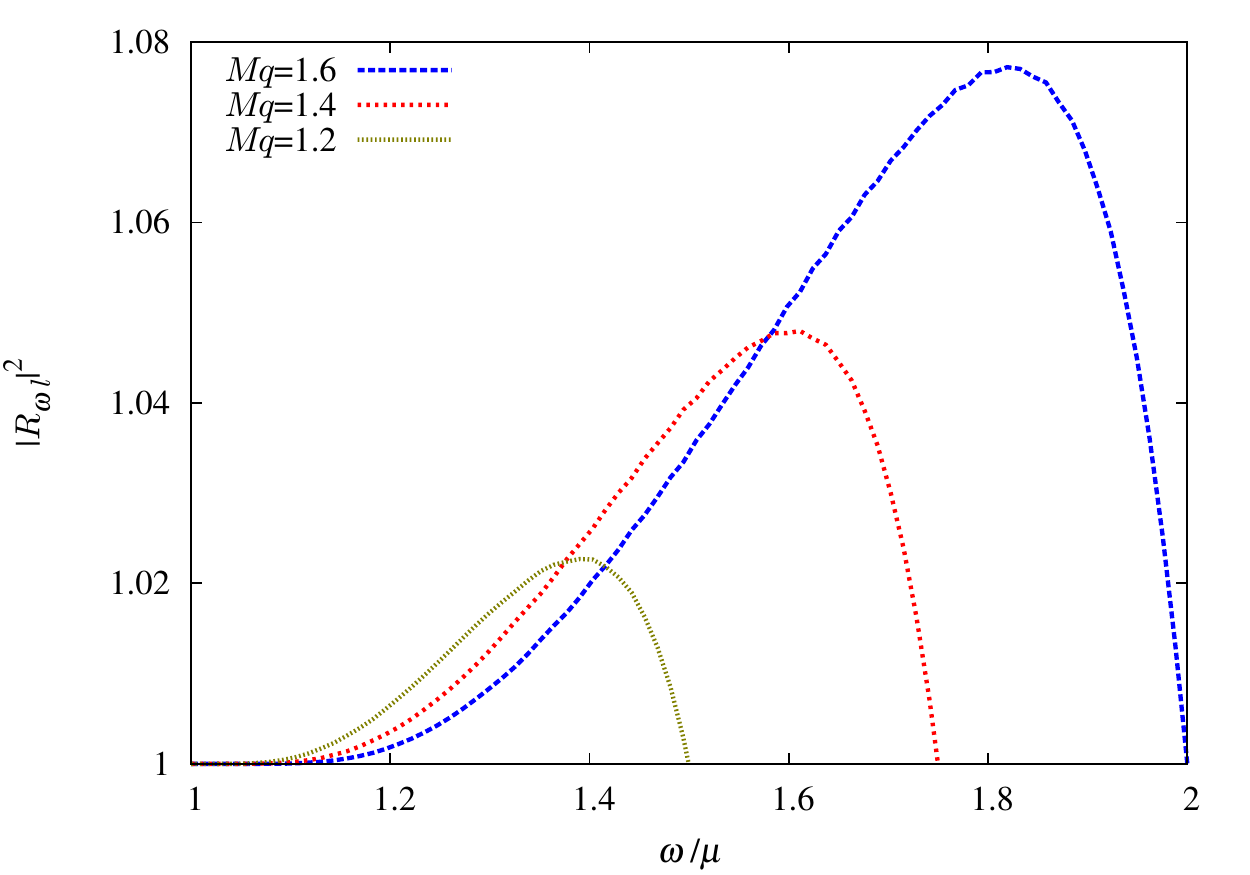}
\caption{Reflection coefficient as a function of the frequency for $M\mu=0.4$, $Q/M=0.8$, $l=0$ and different choices of $Mq$.}
\label{rcq}
\ef

\bf
\includegraphics[width=\columnwidth]{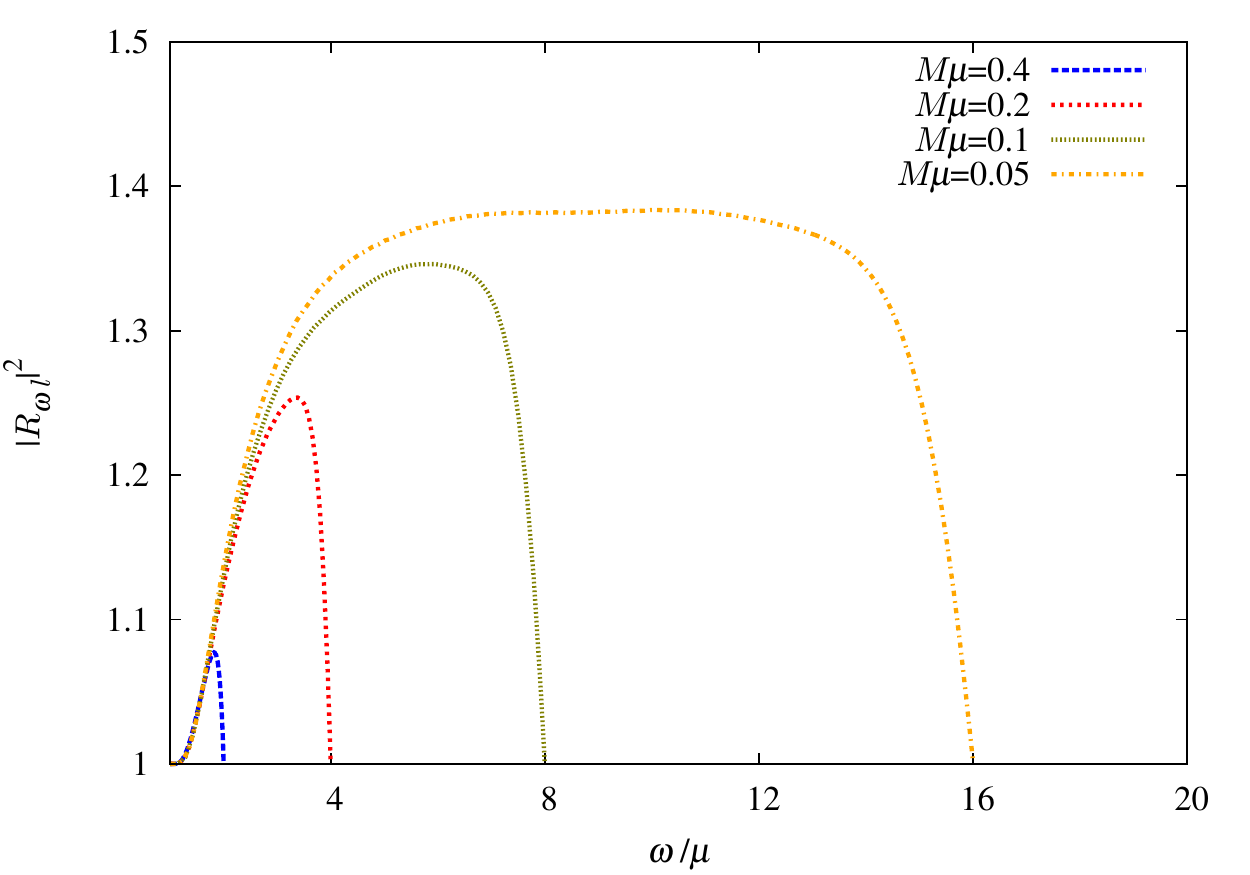}
\caption{Reflection coefficient for $Q/M=0.8$, $M q=1.6$ and different values of $M\mu$. As in Fig. \ref{rcq}, here we only exhibit the values of $|R_{\omega l}|^2$ which are greater than 1.}
\label{rcm}
\ef

\bf
\includegraphics[width=\columnwidth]{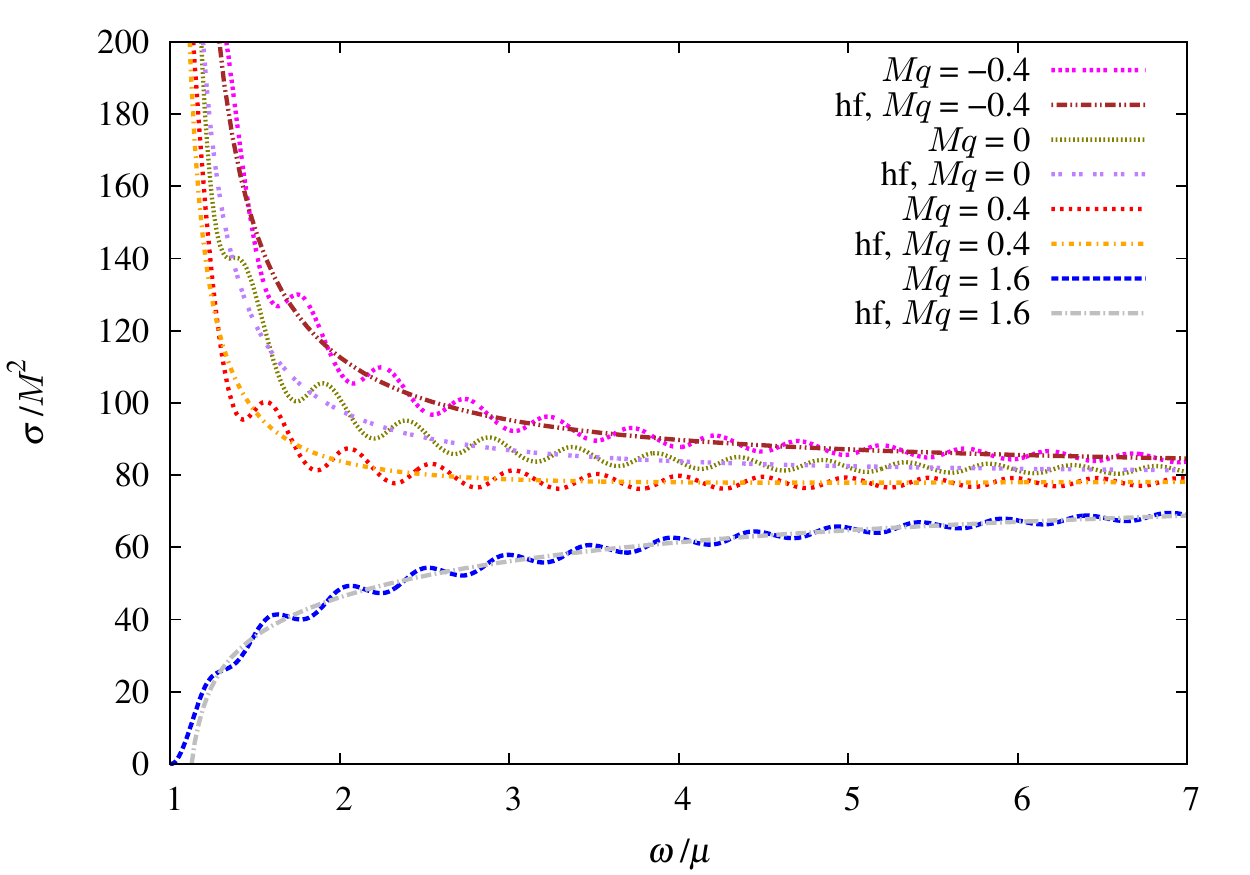}
\caption{Total absorption cross section as a function of the frequency, for $M\mu=0.4$, $Q/M=0.4$ and different choices of $Mq$.}
\label{tac}
\ef

\bf
\includegraphics[width=\columnwidth]{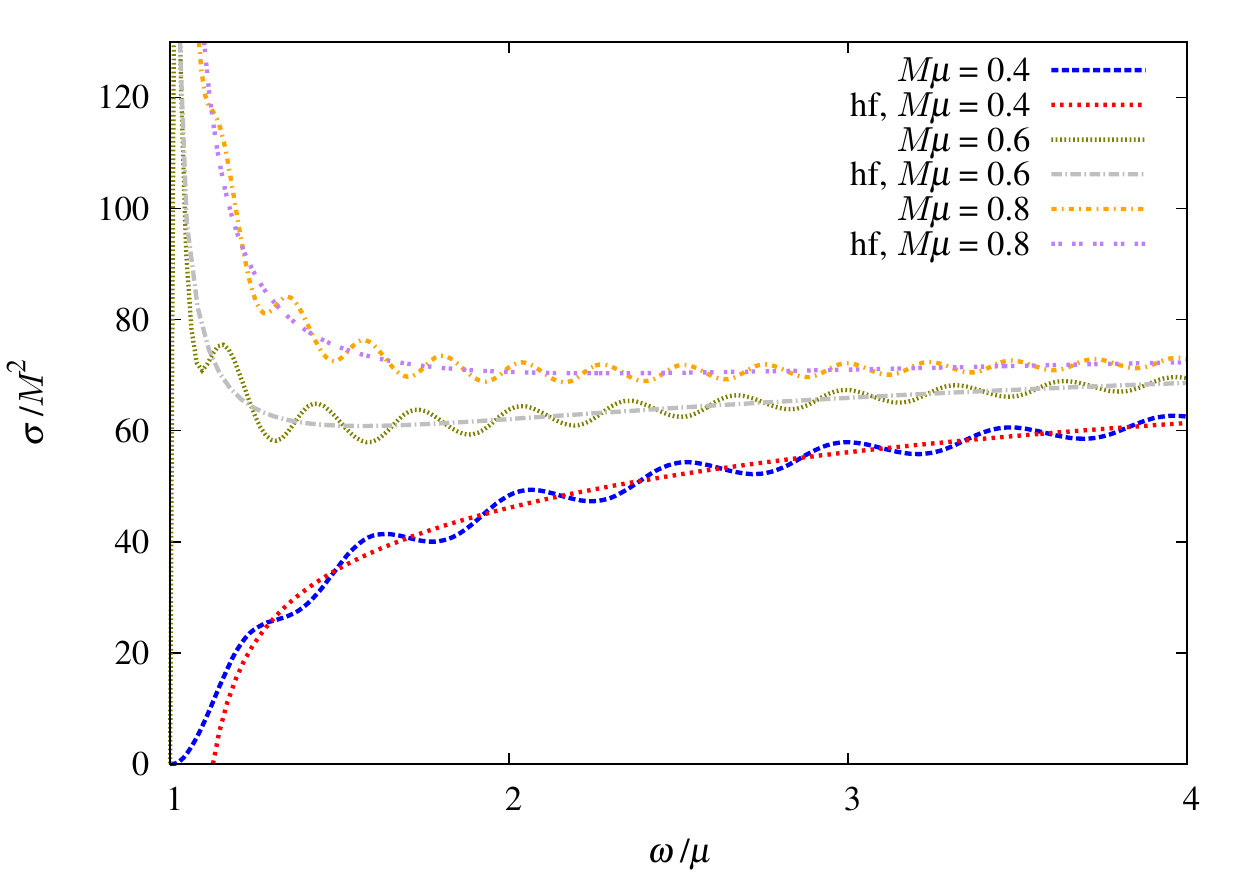}
\caption{Total absorption cross section for $Q/M=0.4$, $Mq=1.6$ and different values of $M\mu$. As in Fig. \ref{tac}, here we also exhibit the corresponding high-frequency result.}
\label{tacm}
\ef

\bf
\includegraphics[width=\columnwidth]{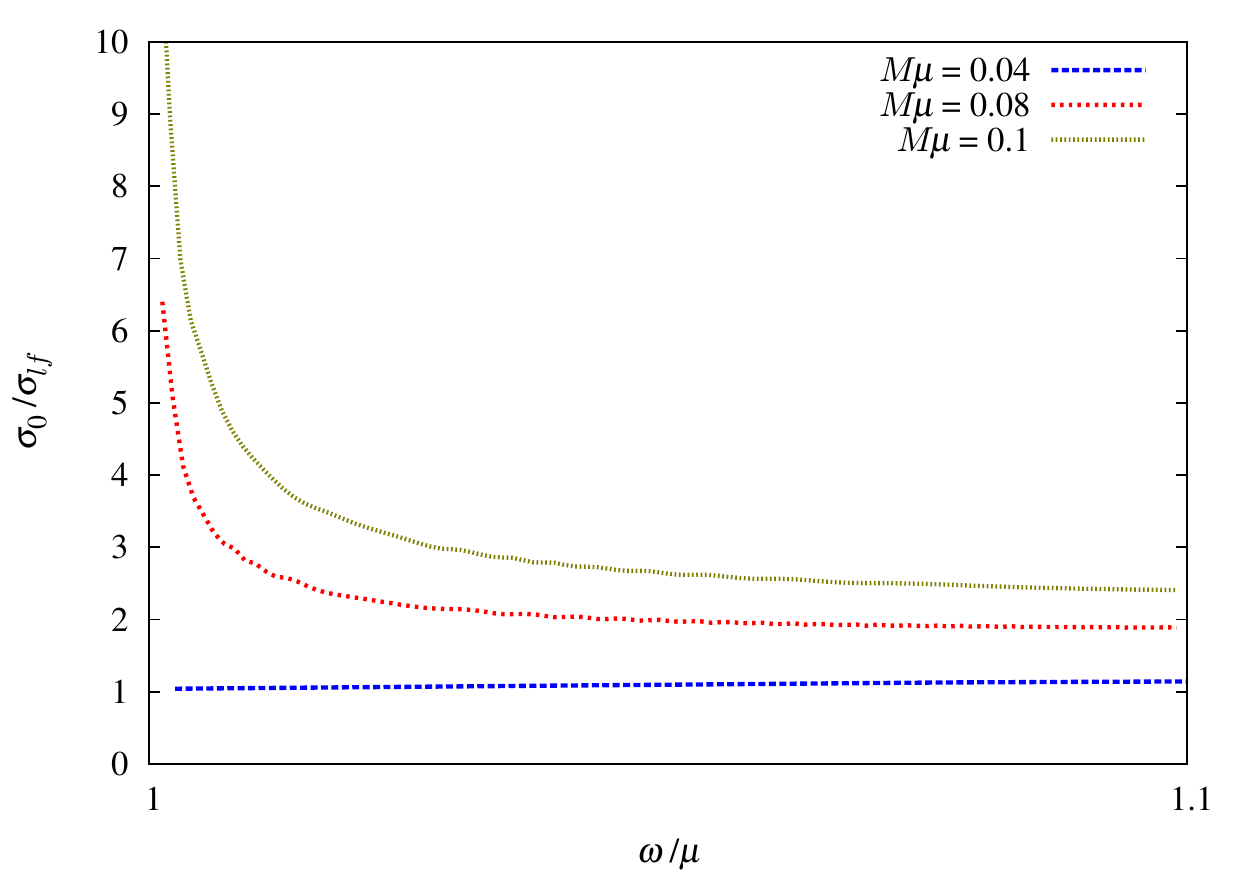}
\caption{Comparison between the partial absorption cross section for $l=0$ ($\sigma_0$) obtained numerically, and the low-frequency approximation ($\sigma_\textit{lf}$) for different choices of $M\mu$. We have chosen $Q/M=0.4$ and $Mq=0.1$ }
\label{pacl}
\ef

\bf
\includegraphics[width=\columnwidth]{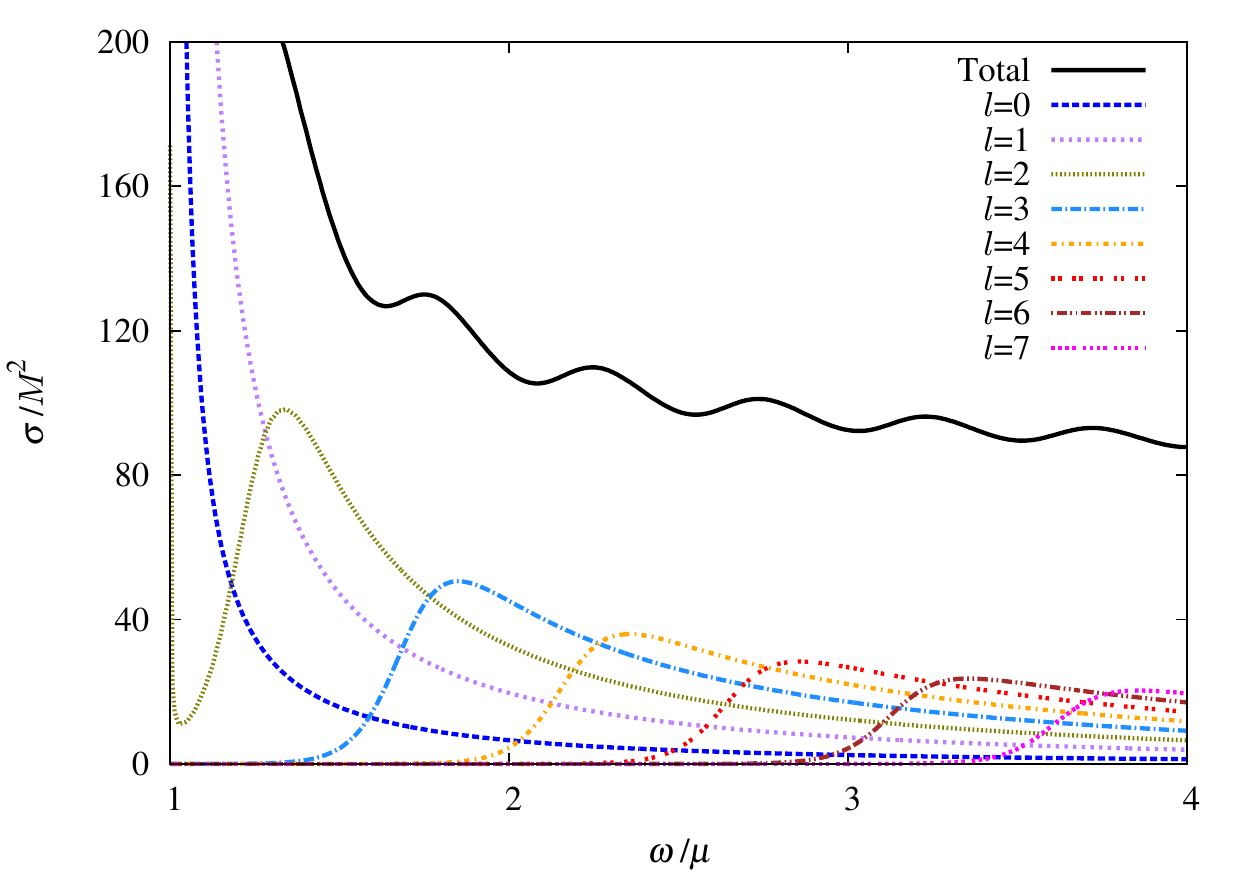}
\caption{Partial and total absorption cross sections for $M\mu=0.4$, $Q/M=0.4$ and $Mq=-0.4$.}
\label{tpa2}
\ef

\bf
\includegraphics[width=\columnwidth]{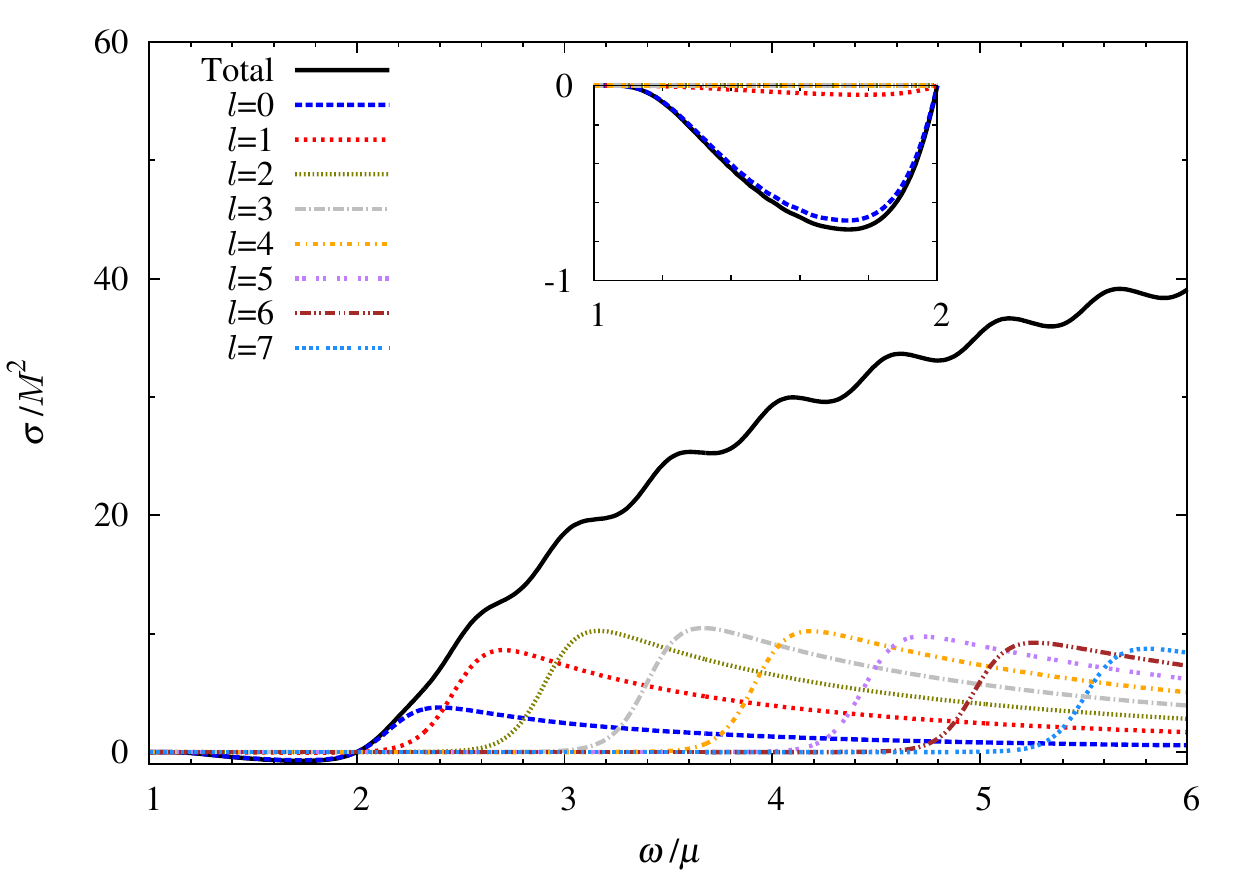}
\caption{Partial and total absorption cross sections for $M\mu=0.4$, $Q/M=0.8$ and $Mq=1.6$.}
\label{tpa}
\ef

\section{Discussions and final remarks}
\label{fr}
We studied the Klein-Gordon equation for a charged and massive scalar field in a Reissner-Nordström background. We found the reflection coefficient and the absorption cross section, comparing numerical and approximate analytical results. Some of the features are similar to the chargeless scalar case, but the charge introduces new aspects.

We obtained that the total absorption cross section oscillates around the high-frequency limit and exhibits a wiggly pattern as a consequence of the individual contributions of the partial absorption cross sections, which present maxima for different values of the frequency. Another interesting feature observed is that as we increase the charge coupling $qQ$, the absorption cross section gets smaller. This is due to the presence of a repulsive electromagnetic interaction (the Lorentz force) for $qQ>0$ competing with the gravitational interaction, causing the decrease of the absorption. 

If the charges of the field ($q$) and of the black hole ($Q$) have opposite signs, the absorption cross section is bigger than for the chargeless field case, due to the Lorentz attraction between the charges.

It is also worth emphasizing that for $Qq>0$, in the limit $\omega \rightarrow \mu$, we can have finite values for the absorption cross section of the charged massive field; while for the case in which $q=0$ the total absorption cross section always tends to infinity in this limit. Therefore the Lorentz repulsion force can render finite the low-frequency limit of the absorption cross section.

Concerning the analytical results, for the high-frequency limit we have been able to find the geometric optics limit of the absorption cross section and the numerical results oscillate around this limit. A similar behaviour is observed in the absorption of the chargeless massive scalar field by a Reissner-Nordström black hole. For the low-frequency limit we found an analytical approximation for the absorption cross section in the case when $M\mu$ and $Qq/r_+$ are very small. The result we found can be regarded as a generalization of the one obtained for the chargeless massive scalar field \cite{Benone:2014qaa}.

We have shown that we can have superradiance of a charged and massive scalar field in a Reissner-Nordström background and that this superradiance increases with the increase of $Qq$ and with the decrease of $M\mu$. As we have seen, both the partial and the total absorption cross section can be negative. Partial waves are associated to spherical waves which, when appropriately summed, give rise to a planar wave. Our results show that planar scalar waves can be superradiantly amplified by black holes. This contrasts with the case of the rotating Kerr black hole \cite{Macedo:2013afa}, for which we also have negative partial scalar absorption cross sections, but they combine in a way that the total scalar absorption cross section is always positive. Even for the acoustic black hole analogue system with rotation, namely the draining bathtub, superradiance in not enough to imply in a negative total absorption cross section \cite{Oliveira:2010zzb}.

We can understand this difference between the uncharged rotating and charged black hole cases by realising that for the rotating case we have, for the total scalar absorption cross section, besides the sum in $l$, a sum in $m$. Consequently, when the partial contributions of $m$ are added, a positive sum is always obtained, resulting that planar scalar waves impinging in a Kerr black hole never present negative total absorption cross section.

Superradiant amplification of planar waves has also been reported for optical systems, related to electromagnetic scattering in active media \cite{Alexopoulos:78,Kerker:78,Kerker:79}.

\section{Acknowledgments}
The authors would like to thank Conselho Nacional de Desenvolvimento Cient\'ifico e Tecnol\'ogico (CNPq), Coordena\c{c}\~ao de Aperfei\c{c}oamento de Pessoal de N\'ivel Superior (CAPES), Funda\c{c}\~ao Amaz\^onia de Amparo a Estudos e Pesquisas do Par\'a (FAPESPA) and Marie Curie action NRHEP-295189- FP7-PEOPLE-2011-IRSES for partial financial support. The authors are also grateful to Caio F. B. Macedo for discussions and suggestions.

\bibliography{rn_abs}

\end{document}